\documentstyle[12pt, aasms4, epsfig]{article}
\catcode`@=11
\def\eqalign#1{\null\,\vcenter{\openup\jot\m@th
  \ialign{\strut\hfil$\displaystyle{##}$&$\displaystyle{{}##}$\hfil
      \crcr#1\crcr}}\,}
\def\eqalignleft#1{\null\,\vcenter{\openup\jot\m@th
  \ialign{\strut$\displaystyle{##}$\hfil&$\displaystyle{{}##}$\hfil
      \crcr#1\crcr}}\,}
\catcode`@=12

\def\lax    {\ifmmode{_<\atop^{\sim}}\else{${_<\atop^{\sim}}$}\fi}
\def\gax    {\ifmmode{_>\atop^{\sim}}\else{${_>\atop^{\sim}}$}\fi}
\def\kms    {\ifmmode{{\rm ~km~s}^{-1}}\else{~km~s$^{-1}$}\fi}

\def\bk{\lower 6pt\hbox{${\buildrel k\over \sim}$}}
\def\bv{\lower 6pt\hbox{${\buildrel v\over \sim}$}}

\begin{document}
\textwidth 6.5truein
\textheight 9.25truein
\topmargin -1.5cm
\title{\bf 
The Large Scale Structures in the Solar System:\\
II. Resonant  Dust Belts\\
Associated With the Orbits of Four Giant Planets
}
\author{\bf Nikolai N. Gor'kavyi\altaffilmark{1}}
\affil{
Laboratory for Astronomy and Solar Physics, NASA/Goddard Space Flight 
Center\\ Greenbelt, MD 20771; also 
Simeiz Department, Crimean Astrophysical Observatory, Simeiz 334242, Ukraine
}
\altaffiltext{1}{NRC/NAS Senior Research Associate; e-mail: 
gorkavyi@stars.gsfc.nasa.gov}
\author{\bf Leonid M. Ozernoy\altaffilmark{2}}  
\affil{5C3, Computational Sciences  Institute and Department of Physics 
\& Astronomy,\\ George Mason U., Fairfax, VA 22030-4444; also Laboratory for 
Astronomy and Solar\\ Physics, NASA/Goddard Space Flight Center, Greenbelt, 
MD 20771} 
\altaffiltext{2}{e-mail: ozernoy@science.gmu.edu; ozernoy@stars.gsfc.nasa.gov}
\author{\bf Tatiana Taidakova\altaffilmark{3}}
\affil{
Simeiz Department, Crimean Astrophysical Observatory, Simeiz 334242, Ukraine
}
\altaffiltext{3}{e-mail: gorkavyi@stars.gsfc.nasa.gov}
\bigskip

\begin{abstract}
In part I, using an effective computational approach, we have reconstructed 
the population of dust sources between Jupiter and Neptune. Here, in
part II, we present the results on distribution of dust produced by
157 real sources (100 Jupiter-family comets with semi-major axes $a<20$ AU, 
 51 Kuiper belt, and 6 Centaur objects) as well as 211 fictitious sources 
taken from our computed sample. The following processes that influence
the dust particle dynamics are taken into account: 1) gravitational 
scattering on four giant planets; 2) planetary resonances; and 3) the 
Poynting-Robertson (P-R) and solar wind  drags. A file consisting of 
$0.9\times 10^6$ particle positions has been computed to simulate 
the dust distribution in the outer parts of the Solar system.
We find that this distribution is highly non-uniform, with most of the dust 
concentrating into four belts associated with the orbits of four giant 
planets, with a sharp rise (depending on the size of particles) at the 
innermost part of the ring. As 
in I, we reveal a rich and sophisticated resonant structure of these belts
containing families of resonances and gaps. A dissipative nature of the P-R
drag results in specific features of particle's capture into, and evolution
in, the resonances.

Based on our simulations, we expect a new, quasi-stationary dust
population to exist in the belts near Jupiter and Saturn, which is highly
inclined and possesses large eccentricities. This population is basically
 non-resonant and is an important addition to otherwise resonant dust belts.

The simulated dust is likely the main source of the zodiacal light in 
the outer Solar system, which will be analyzed in our further work. 

\end{abstract}
\newpage
\section*{1. Introduction}
As shown in part I of our work (Ozernoy, Gorkavyi, \& Taidakova 
1998, referred hereinafter as I), 
comets coming from trans-Neptune regions (mostly from the Kuiper belt)
are scattered gravitationally on all giant planets and form
a quasi-stationary  population of sources between Jupiter and Neptune. This 
paper aims at examining the distribution of dust produced by those sources.
The dynamics of this dust is determined by three main effects: (i) 
the Poynting-Robertson (P-R) drag (including radiation pressure and 
solar wind drag), (ii) gravitational scattering on the 
four giant planets, and (iii)~resonances with those planets. 

An extensive work on dust particle evolution governed 
by the above effects has been done by a number of
investigators (Weidenschilling \& Jackson 1993, Roques et al. 1994, 
Lazzaro et al. 1994, Liou \& Zook 1997,
Gor'kavyi, Ozernoy, Mather, \& Taidakova 1997, Kortenkamp \& Dermott 1998).
The present paper makes a next step by accounting for the following 
new elements: (i) as sources of dust, we use all known Kuiper belt bodies
and, as additional sources, we use fictitious minor bodies from all 
cometary-asteroidal belts of the four giant planets computed in I; and
(ii) after computing a stationary distribution of dust particles
in the space of orbital elements, $n(a,e,i)$, we employ an analytical method
to derive the 3-D model of the interplanetary dust cloud 
in the outer Solar system. 

In Section 2, we discuss the sources of dust particles in the outer 
Solar system. Sec.~3 describes our numerical method that enables us,
in conjunction with an analytical approach, to  compute the 3-D
distribution of dust in the outer Solar system. Sec.~4 contains
the results of these computations, which reveal the global dust distribution 
as well as interesting details of its resonant structure. Our conclusions
are presented in Sec.~5.

\vskip 0.15truein
\section*{2. The Sources of Dust Particles}

There is a mounting evidence that the sources of the interplanetary dust 
particles (IDPs) cannot be entirely reduced simply to 
those comets which produce the observed dust tails or/and to asteroids
which are thought to be responsible for the observed `dust bands' 
in the IDP emission -- a number of facts forces to suspect
that additional sources of the interplanetary dust must exist. 
Among others, two such facts are worth mentioning:
(i) According to Pioneer's 10 and 11 data, 
the dust particles are seen up to 18 AU (Humes 1980; Divine 1993), which
implies the existence of a dust source beyond 4 AU (Flynn 1996); and
(ii) Chemical analyses and other space-based data indicate 
that a part of IDP spent in space a much larger time that the
typical asteroidal and cometary particles, which 
is a strong evidence in favor of other, along with the known comets and
asteroids, sources of dust in the Solar system  (Flynn, 1996).

A new, for a long time neglected factor   
in the problem of the IDP origin
is the Kuiper belt so that a third component 
of the IDP cloud might be the `kuiperoidal' dust (Backman et al. 1995).
In our opinion, the Kuiper belt influences the formation of the
IDP cloud in two ways:
1) as a source of small-size particles slowly drifting toward the
Sun under a combined action of the PR-drag, gravitational scattering,
and  influence of resonances [evolution of 80 such  particles 
was computed by Liou, Zook, \& Dermott (1996)];
and 2) as a source of trans-Jovian comets. 
It is commonly agreed that
the Jupiter-family comets are produced by transporting the comets from
the Kuiper belt via gravitational scattering on the four giant planets when
each planet scatters the comets both toward and away from the Sun (Levison 
\& Duncan 1997). Our numerical simulation 
described in I indicates that, between Jupiter and Neptune, there is 
a numerous population of minor bodies forming four cometary-asteroidal
belts near the orbits of all giant planets. 

The minor body families of Saturn, Uranus, and Neptune should contain
progressively  larger numbers of comets than one sees  near
Jupiter. Even despite a many-fold decrease of the solar heat intensity
at such large distances, those numerous comets may produce dust
in amounts comparable to that from a few active J-comets.
Complementary mechanisms of dust release from kuiperoids and Centaurs
between Jupiter and Neptune can include
impacts of large grains and the  solar wind. We refer to observational
data indicating,  for a number of kuiperoids 
and Centaurs,  a steady cometary activity for years 
(e.g. Brown \& Luu, 1998 and refs. therein).

The kuiperoidal dust experiences the same dynamical effects as the
asteroidal dust, with the only difference
that, due to a slower PR-drift and a stronger influence of the giant
planets, the role of gravitational scattering and resonance captures 
must be more important for it. In what follows, while computing the dynamics
of kuiperoidal dust, our model incorporates 5 dust components of different 
origin associated with respected belts of minor bodies, viz., 
a) Kuiper belt; b) Neptunian belt; c) Uranian belt; d) Saturnian belt; 
and e) Jovian belt. We  make use of the available list of 51
kuiperoids with known orbital parameters as well as 6 Centaurs
(Marsden 1998). In order to reduce to a minimum the influence of poorly
known observational selection effects, we add a sample  
of 211 fictitious sources randomly taken from our simulation of the minor
body population between Jupiter and Neptune. The distribution of the
used dust sources in the orbital coordinates $a,e,i$ is shown in Fig.~1.

\section*{3. Computational Method: Simulation of a Quasi-stationary 
Distribution of Dust Particles in the Outer Solar System}

We calculate the orbital elements $a,e,i$ of particles  starting from a 
source of dust and then drifting toward the Sun under the P-R drag.
On its way to the Sun, the particle undergoes the gravitational influence
of the four giant planets. 
As in I, we adopt the approximation of a restricted
3-body problem (the Sun, the planet on a circular orbit, and a massless 
particle). In order to reduce the computations to a restricted
3-body problem, we assume that the particle, while being in the planet's 
zone of influence, does not feel gravitational perturbations from the three 
other giant planets.
The planet's zone of gravitational influence  
in the $(a,e)$-plane of orbital coordinates is defined by:
$$a(1-e)\leq a_p ~~~{\rm if}~ a>a_p,$$
$$a(1+e)\geq a_p ~~~{\rm if}~ a<a_p,$$
where $a$ is the semi-major axis of a test body,
$a_p$ is the semi-major axis of the planet, and
$e$ is eccentricity of the test body.
The above approximation is only temporary and will be abandoned in our 
further work. 

As a convenient approach to simulate a quasi-stationary distribution 
of dust particles, we applied the following computational procedure: 
a record of particle's orbital elements 
was taken after certain number of revolutions (usually each 
10 revolutions) of the planet
around the Sun and these data were then used to characterize the positions 
of {\it many} particles over the entire time span, beginning from
an initial instant till the instant of particle's death (impact on planet, 
the Sun, or particle's ejection from the Solar system).

In order to explore the contribution of each planet's
cometary belt into the general dust distribution, we normalize to 
unity the dust production in each cometary belt. In physical terms, it 
implies that a smaller minor body abundance in the innermost giant planets 
is compensated by a larger dust production due to their proximity to the Sun.
In our further work, we hope to abandon this approximation by an accurate
computing the `transfer function' (a fraction of minor bodies
gravitationally scattered by all outer planets into the given planet's zone 
of influence) as well as by a reasonable estimation of the cometary
activity as a function of the comet's orbital elements.

We computed more than 360 stationary distributions $n(a,e,i)$ of dust 
particles from 51 kuiperoids and 211 fictitious sources. These distributions
form a file consisting of $0.9\times 10^6$ positions. Numerical integrator 
described in Taidakova (1997) and  Taidakova \& Gor'kavyi (1999) was 
employed. Details of computational runs are given in Table 1.
\section*{4. The Results: Four Dust Belts and Their Resonant Structure}

\subsection*{4.1. Four Dust Belts}

Our computations have been done for the P-R parameter (the radiation
pressure to gravitational force ratio) $\beta=0.1$ 
and the solar wind drag to P-R-drag ratio $=$0.35 (Gustafson 1994).
For $\beta=0.1$, resonances computed in I are shifted by a factor 
$(1-\beta)^{1/3}=0.965$. The larger the value of $\beta$, the larger is 
the drift velocity and the smaller is the probability of a resonant capture.

Representative results of the orbit integrations are shown, in the orbital
coordinates $a,e$ and $a,i$, for Neptune in Fig.~2a,b
and for Jupiter in Fig.~3a,b.  
Capture of {\it kuiperoidal} particles 
into Neptune's dust belt occurs predominantly into 3:2 resonance, 
which takes the major responsibility for the entire dust belt.
This might be partly explained by observational selection (just a few sources 
is presently known beyond 2:1 resonance). 
In   Jupiter's zone of influence, high-eccentricity cometary particles 
 are  mostly captured into two resonances,
viz., 3:2 and 1:1. The capture into 1:1 resonance for asteroidal 
dust particles with $\beta=0.26$ was considered by Liou \& Zook (1995).  


The general picture of dust distribution in the outer part of the 
Solar system obtained by summation of the computed particle
distribution functions for every giant planet's zone of influence
is shown in Fig.~4. 

We find that the simulated dust distribution is highly non-uniform, 
with most of the dust concentrating into four belts near the orbits 
of four giant planets. 

The major part of the simulated Neptune's dust belt is located between 
24 AU and 60 AU and forms a flat dense disk.
The simulated Uranian, Saturnian, and   Jovian dust belts are essentially 
overlapped and form sophisticated dust structures which, in their central
parts, are less dense compared to Neptunian dust belt.

The most remarkable feature found in our simulations is that 
they indicate the existence of a new quasi-stationary, highly inclined
dust population with pericenters near Jupiter and Saturn. 
In Figs.~4a,b, this population is seen as a `Chinese wall'-like structure. 
This structure is found to be more steep and of a larger density for 
larger-size particles such as $d=12\mu$m ($\beta=0.037$).
The above-mentioned
quasi-stationarity results from a balance between the tendencies for 
particle's semi-major axis $a$ and eccentricity $e$ to increase due to 
gravitational scattering on the planet and to decrease due to the P-R drag. 
As for particle's inclinations $i$, they substantially increase due 
to gravitational scattering and influence of resonances. 

We note in passion that  the `Chinese wall'-like
features of dust distribution near the giant planets, 
especially as massive as Jupiter, can serve as signatures of exo-planets
in the circumstellar disks.

Distribution of dust density in the ecliptic plane, which is of obvious
practical interest for current and future spacecraft missions,
is shown in Figs.~5 and 6. It reveals substantial rises and falls in
dust number density. 
The most remarkable result is a rather steep rise (somewhat 
depending on the particle's size) of dust density in the innermost
part of all dust belt, especially of Jupiter's (at $R=4.3$ AU) and Neptune's
(at $R=24$ AU).
Rather sharp inner edges and the respected steps in dust density 
distribution are expected to characterize each giant planet's  dusty belt
at 0.85$a_{\rm planet}$.
Neptune's dust belt is expected to have
the largest number density of particles in the ecliptic plane. 

We note that the ecliptic dust density is rather sensitive to a contribution
of low-inclination particles [$n\propto (\sin i)^{-1}$]. A two-peak structure
between Jupiter's and Saturn's orbit (see Fig.~5a) is produced by the dust 
from Neptune's zone; this structure is formed by low-inclination 
particles ($i<1^\circ$) captured into 3:2 resonance with Jupiter. With an
improved statistics, the heights of those peaks may change. However, the
two-peak structure in Neptune's zone, which has been simulated  by a large
number of particles, is robust.

The general distribution of dust in the outer Solar system, as follows
from our simulation, is shown in Figs.~7 and 8.

\subsection*{4.2. Resonant Structure of the Four Belts}
Just as the parent cometary populations, the simulated dust belts
reveal a sophisticated resonant structure containing rich 
families of resonances and gaps. The main difference with the 
parent sources is that the resonant capture of dust particles occurs in a
{\it dissipative} way, and this process takes place  
both inside and outside the zone of planet's influence.

The distribution of interior resonances in the $a,e,i$-space is
characterized by the presence of numerous gaps. As for the exterior 
resonances, the particles are dissipatively captured into those resonances 
(usually outside the triangle zone). Subsequent 
evolution is characterized by an increase of eccentricity, with 
oscillations in $e$ and $i$, whose amplitude is especially large where both 
$e$ and $i$ are high enough.
  
 In a resonance $(j+1)/j$, while eccentricity is close to the maximal one, 
$e_{\rm max}=\sqrt {0.4/(j+1)}$ (Weidenschilling \& Jackson 1993),
the particle's life time  is expected to be long  enough.
This results in two-peak structures seen in Figs.~5 and 6. The characteristic 
shape of such structures follows from the Kepler motion laws (Kessler 1981,
Gorkavyi et al. 1997b). The inner and outer edges of each structure are
given by $a_{res}(1-e_{\rm max})$ and  $a_{res}(1+e_{\rm max})$, 
respectively. For all resonances, in a good approximation, the position
of the inner edge is $\approx 0.85a_{\rm planet}$. At this position, we expect
to find rather sharp and steep inner edges and the respected steps in 
the dust density distribution (all that somewhat 
depending on the particle's size) for each giant planet's  dusty belt.

Finding, by dust detectors on spacecraft,  of such dust density peaks 
expected  at  $R\approx4.3$ AU in
Jupiter's  zone and at $R\approx24$ AU in  Neptune's  zone 
would be a direct confirmation of  the dust belts, as well as their
 resonant nature, 
simulated in the present work.

The resonant nature of the simulated dust belts is seen in Fig.~7 :
the major part of Neptune's dust belt has an azimuthal asymmetry. The latter 
is revealed as an enhanced concentration of dust particles in Neptune's
trailing zone.

\section*{5. Conclusions}

1. We find that the simulated dust distribution is highly non-uniform, 
with most of the dust concentrating into four belts near the orbits 
of four giant planets. Those belts would be a challenging
target to discover by space missions, either ongoing ({\it CASSINI}) or
forthcoming ({\it STARDUST, ISAS PLANET B}).

2. Our simulations indicate the existence of a new quasi-stationary, highly
inclined dust population with pericenters near Jupiter and Saturn. This 
quasi-stationarity results from a balance between the tendencies for 
particle's semi-major axis $a$ and eccentricity $e$ to increase due to 
gravitational scattering on the planet and to decrease due to the P-R drag. 
As for particle's inclinations $i$, they substantially increase due 
to gravitational perturbations from Jupiter and Saturn. This highly inclined 
population produces a very large, high altitude `wall' (see Fig.~4). 

3. Just as the parent cometary populations, the simulated dust belts
reveal a sophisticated resonant structure containing rich 
families of resonances and gaps. The main difference with the 
parent sources is that the resonant capture of dust particles occurs in a
{\it dissipative} way, and this process takes place  
both inside and outside the zone of planet's influence.

4. The distribution of interior resonances in the $a,e,i$-space is
characterized by the presence of numerous gaps. As for the exterior 
resonances, the particles are dissipatively captured
into those resonances (usually outside the triangle zone). Subsequent 
evolution is characterized by an increase of eccentricity, with 
oscillations in 
$e$ and $i$ (the amplitude of these oscillations is especially large 
where both $e$ and $i$ are high enough).

5.  A rather long life time in a resonance, while eccentricity is close to  
the maximal one, results in a rather steep rise (somewhat 
depending on the particle's size) of dust density in the innermost
part of all dust belts, especially of Jupiter's (at $R=4.3$ AU) and Neptune's
(at $R=24$ AU).
Rather sharp inner edges and the respected `steps' in the dust density 
distribution are expected to characterize each giant planet's  dusty belt
at 0.85$a_{\rm planet}$.
Neptune's dust belt is expected to have both 
the largest `step' and number density of particles in the ecliptic plane 
(Fig.~5). 

6. The simulated dust is likely to be the main source of the zodiacal light 
emission in 
the outer Solar system, which will be analysed in more detail in our further
work. 

7. The revealed features of dust distribution near the giant planets, 
especially as massive as Jupiter, can serve as signatures of exo-planets
in the circumstellar disks.


\vspace{0.1in}
{\it Acknowledgements.} This work has been supported by NASA Grant NAG5-7065 
to George Mason University. N.G. acknowledges the NRC/NAS associateship.
T.T. is thankful to the American Astronomical Society for a Small Research
Grant from the Gaposchkin's Research Fund.

\newpage

\centerline{TABLE 1}
\medskip
\hrule
\medskip
\centerline{Details of Computational Runs}
\medskip
\hrule
\smallskip
\hrule
\medskip
\centerline{\vbox{
\halign{ \hfil # \hfil  & \hfil # \hfil  & \hfil # \hfil 
& \hfil # \hfil  & \hfil # \hfil  & \hfil # \hfil   
& \hfil # \hfil
\cr
Source of dust  & Neptunian & Uranian & Saturnian & Jovian & kuiperoids   
& J-family \cr
\hfil  & cometary &  cometary  & cometary  & cometary & \& Centaurs &comets 
\cr
\hfil  & belt & belt & belt & belt &  & \cr
\noalign{\vskip 10pt}
 Number of sources \hfil  & 59 & 50 & 55 & 47 & 57 & 100 \cr
\noalign{\vskip 10pt}
 Typical life time \hfil & from $1\times 10^6$     &from $2\times 10^5$   &
 from $5\times 10^4$  &  from $2\times 10^4$  &  from $1\times 10^6$   &
 from $1\times 10^4$  \cr
of particles, yrs $(^1)$  \hfil & to $3\times 10^7$    &  to 
$1.5\times 10^7$ &  to $3\times 10^6$ &  to $1\times 10^6$&  to $3\times 10^7$&
  to $1\times 10^6$ \cr  
\noalign{\vskip 10pt}
 Number of computed \hfil   &     &  &  & & & \cr
positions$(^2)$\hfil& 245,543 & 191,463 & 131,509 & 53,454 &249,520 & 26,939\cr
\noalign{\vskip 10pt}
 Number of positions \hfil  &  &  &  &  \cr
used for modelling$(^3)$\hfil & 24,554 & 19,146 & 13,150
& 5,345 & 24,952 &--\cr
\noalign{\vskip 10pt}
Number of positions \hfil  &  &  &  &  \cr
plotted in Figs. 2 to 3\hfil & -- & --  & -- 
& -- & 31,575 $(^4)$ & 26,939\cr
\noalign{\vskip 8pt}
}}}
\hrule
\bigskip

$(^1)$ until the particle impacts the Sun or is ejected 

$(^2)$ taken with time step $=$ 10 revolutions of the planet.

$(^3)$ every $1/10$ position from the previous line.

$(^4)$ taken with time  step $=$ 30 revolutions of the planet.
\newpage
\centerline{\bf Figure Captions}
{\bf Figure 1}.
 
{\bf a}.
Distribution of minor bodies as the sources of dust the between orbits of 
Jupiter and Neptune shown, along with asteroids of the mail belt and Centaur 
objects, on the $(a,e)$-plane.
Crosses stand for asteroids of the main belt (100 objects), 
triangles stand for Jupiter-family comets (112 objects), 
squares stand for Centaurs (6 objects), diamonds stand for kuiperoids 
(50 objects), and stars stand for fictitious bodies taken from our
simulations (paper I). 
The dot-dashed line separates 
the comets with pericenters less than 2 AU  (located above the line) from 
yet unrevealed comets having larger pericenters.
The triangle zones of each giant planet's 
gravitational influence are shown by dashed lines.

{\bf b}. Same minor bodies as in {\bf a} shown on the $(a,i)$-plane.

\vspace{0.1in}
{\bf Figure 2}.

{\bf a}. Distribution of dust particles near Neptune's orbit produced
from 50 kuiperoids.
The Neptune's triangle zone is shown by heavy lines. For illustration 
purposes, gravitational influence of three other giant planets upon 
the distribution of dust is neglected.
Positions of appropriate exterior and interior resonances are shown by 
arrows.

{\bf b}. Same as in {\bf a} shown in $(a,i)$-coordinates.  

\vspace{0.1in}
{\bf Figure 3}.

{\bf a}. Distribution of dust particles near Jupiter's orbit produced
by 100 comets of Jupiter family with $a<20$ AU.
The  Jupiter's triangle zone is shown by heavy lines. For illustration 
purposes, gravitational influence of three other giant planets upon 
the distribution of dust is neglected.
Positions of appropriate exterior and interior resonances are shown by 
arrows.

{\bf b}. Same as in {\bf a} shown in $(a,i)$-coordinates.  

\vspace{0.1in}
{\bf Figure 4}. 
{\bf a}. Section of the simulated dust distribution in the 
plane perpendicular to the ecliptic plane  within 30 AU .
Adjoining regions of different colors have density contrast 10:1 
(the densest regions are in the ecliptic plane).

{\bf b}. The same within 80 AU. 

\vspace{0.1in}
{\bf Figure 5}. Dust number density in the ecliptic plane (normalized to 
the maximum density of kuiperoidal dust, $n_{\rm max}$) as a function
of heliocentric distance. Computations were performed for $R\geq 2$ AU. 
Within $2<R<4$ AU where influence of Jupiter can be neglected, the dust
number density follows a $R^{-1}$ dependence, as it should be under the 
P-R drag influence. 

{\bf a}. Distribution of kuiperoidal dust (i.e. the dust originated 
in the Kuiper belt 
and then transported inward) and Neptunian dust (i.e. the dust originated
in the Neptune's cometary belt and then transported inward) within 
heliocentric distances of 80 AU.

\vspace{0.1in}
{\bf Figure 6}. Distribution of Uranian, Saturnian, and Jovian dust (i.e. the
dust originated in the respected cometary belts and then transported inward) 
within heliocentric distances of 20 AU.

\vspace{0.1in}
{\bf Figure 7}.
The large-scale structure of the dust cloud in the outer part of the Solar 
system shown face-on. For convenience, positions
of Jupiter, Saturn, Uranus,  and Neptune are indicated. 
Dust particles produced in Kuiper belt and Neptune zone, Uranus 
zone, Saturn zone, and Jupiter zone are shown by the black, blue, red, and 
violet color, respectively. The particle's positions are given
in the frame rotating with the respected planet (i.e. counter clock-wise);  
therefore the resonant particles form a stationary pattern. Note an 
asymmetry in the Neptune's trailing zone.

\vspace{0.1in}
{\bf Figure 8}.
The large-scale structure of the dust cloud in the outer part of the Solar 
system shown edge-on. For convenience, positions
of Jupiter, Saturn, Uranus,  and Neptune are indicated. 
Dust particles produced in Kuiper belt and Neptune zone, Uranus 
zone, Saturn zone, and Jupiter zone are shown by the black, blue, red, and 
violet color, respectively.

\newpage

\centerline{\bf References}
\def\ref#1  {\noindent \hangindent=24.0pt \hangafter=1 {#1} \par}
\smallskip
\ref{Backman, D.E., Dasgupta, A. \& Stencel, R.E. 1995, ApJ 450, L35}
\ref{Divine, N. 1993, J. Geophys. Res. 98E, 17029}
\ref{Flynn, G.J. 1996, in {\it Physics, Chemistry, and Dynamics
  of Interplanetary Dust}, ed. B. Gustafson \& M. Hanner, (San
  Francisco: ASP), ASP Conf. Ser. 104, p. 171}
\ref{Gor'kavyi, N.N., Ozernoy, L.M. \& Mather, J.C. 1997a, ApJ 474, 496}
\ref{Gor'kavyi, N.N., Ozernoy, L.M., Mather, J.C. \& Taidakova, T. 
1997b, ApJ 488, 268}
\ref{ Gor'kavyi, N.N., Ozernoy, L.M., Mather, J.C. \& Taidakova, T.
   1998, Earth, Planets and Space, 50, 539}
\ref{Gor'kavyi, N.N., Ozernoy, L.M., 1999, ApJ  (to be submitted)}
\ref{Gustafson, B.A.S. 1994, Ann. Rev. Earth Planet. Sci. 22, 553}
\ref{Humes, D.H. 1980, J.Geophys.Res.,85, 5841}
\ref{Kessler, D.J. 1981, Icarus 48, 39}
\ref{Kortenkamp, S.J. \& Dermott, S.F. 1998, Icarus 135, 469}
\ref{Lazzaro D.D. et al. 1994, Icarus 108, 59}
\ref{Levison, H.F. \& Duncan M.J. 1997, Icarus 127, 13}
\ref{Liou, J.-C. \& Zook, H.A. 1995, Icarus, 113, 403}
\ref{Liou, J.-C. \& Zook, H.A. 1997, Icarus, 128, 354}
\ref{Liou, J.-C., Zook, H.A. \& Dermott, S.F. 1996, in
{\it Physics, Chemistry, and Dynamics of Interplanetary Dust},
    ed. B. Gustafson \& M. Hanner, (San Francisco: ASP),
ASP Conf. Ser. 104, p. 163}
\ref{Marsden, B.G. 1998, MPEC 1998-v14: Distant Minor Planets}
\ref{Ozernoy, L.M., Gor'kavyi, N.N., and Taidakova, T. 1998, astro-ph/9812479}
\ref{Roques, F. et al. 1994,  Icarus 108, 37}
\ref{Taidakova, T. 1997, in {\it Astronomical Data Analyses,
 Software and Systems VI}, ed. G. Hunt \& H.E.Payne, (San Francisco:
 ASP), ASP Conf. Ser. 125, p. 174}
\ref{Taidakova, T. \& Gor'kavyi, N.N. 1999, {\it
    The Dynamics of Small Bodies in the Solar Systems: A Major Key to
    Solar Systems Studies}, ed. A.E.Roy \& B.A.Steves (in press)}
\ref{Weidenschilling, S.J. \& Jackson, A.A. 1993, Icarus 104, 244.}

\end{document}